\documentclass{PoS}
\usepackage{subfigure}
\usepackage{amsmath,amsfonts}
\usepackage{mathptmx}
\usepackage{mathrsfs}
\usepackage{bbm}
\usepackage{bm}
\usepackage{xfrac}
\newcommand{\beq}{\begin{equation}}
\newcommand{\eeq}{\end{equation}}
\newcommand{\bea}{\begin{eqnarray}}
\newcommand{\eea}{\end{eqnarray}}

\renewcommand{\d}{\delta}
\renewcommand{\l}{\lambda}

\renewcommand{\b}{\beta}
\renewcommand{\a}{\alpha}

\renewcommand{\o}{\omega}

\newcommand{\tk}{\widetilde{k}}

\newcommand{\m}{\mu}
\newcommand{\g}{\gamma}
\newcommand{\bx}{{\mathbf x}}

\newcommand{\s}{\sigma}

\newcommand{\D}{\Delta}

\newcommand{\N}{{\cal N}}

\renewcommand{\th}{\theta}

\newcommand{\oh}{\frac{1}{2}}

\newcommand{\non}{\nonumber}

\newcommand{\rf}[1]{(\ref{#1})}
\newcommand{\ra}{\rightarrow}

\title{Measurements of the Yang-Mills vacuum wavefunctional, and related studies}

\ShortTitle{Measurements of the Yang-Mills vacuum wavefunctional}

\author{\speaker{Jeff Greensite}%
        \thanks{Supported in part by 
the U.S.\ Department of Energy under Grant No.\ DE-FG03-92ER40711.}\\
       Physics and Astronomy Dept., San Francisco State
University, San Francisco, CA~94132, USA\\
       E-mail: \email{greensit@sfsu.edu}}
 
\abstract{I present numerical tests of several recent proposals
for the Yang-Mills vacuum wavefunctional in D=2+1 dimensions.  In these tests the predicted relative values of the
squared wavefunctional, evaluated on a finite set of abelian plane wave and non-abelian constant configurations, are compared to the corresponding values for the true ground state, extracted from Monte Carlo simulations.   
I also study how the 2+1 dimensional color Coulomb potential, averaged over a set of simple trial configurations, varies
with the proximity of those configurations to the Gribov horizon.}

\FullConference{International Workshop on QCD Green's Functions, Confinement and Phenomenology\\
                5-9 September 2011\\
                Trento, Italy}

\begin{document}

\section{Introduction}

   I would like to address two topics in this talk, concerned with Yang-Mills theory and confinement in 2+1 dimensions.  First, there are now a number of different proposals concerning the form of the Yang-Mills vacuum wavefunctional
in $D=2+1$ dimensions, and I will review here some recent numerical tests of these proposals. This is work done in collaboration with  Hrayr Matevosyan, {\v S}tefan Olejn\'{\i}k, Markus Quandt, Hugo Reinhardt,  and Adam Szczepaniak, and is presented in much more detail in ref.\ \cite{Greensite:2011pj}.   My second topic concerns the Gribov horizon in Coulomb gauge.  Specifically, I would like to answer this question:  Suppose we  generate transverse gauge fields arbitrarily close to the first Gribov horizon, drawn from a simple momentum-space probability distribution dictated by the desired  (e.g. Gribov-Zwanziger) form of the gluon propagator.  Are such configurations, by virtue of their proximity to the horizon, sufficient to produce a linear (or near-linear) rise in the color Coulomb potential? 

\section{The Proposed Vacuum Wavefunctionals}

     Since most of the interesting properties of non-abelian gauge theories, such as confinement and chiral symmetry breaking, are actually properties of the vacuum, we might learn more about those phenomena if we knew the explicit
form of the ground state wavefunctional $\Psi_0[A]$.  A very old proposal \cite{Greensite:1979yn} is that, at large scales, the Yang-Mills vacuum wavefunctional looks like
\beq
\Psi_0^{eff}[A] = \N \exp\left[-\oh \m \int d^3 x ~ \mbox{Tr}[F_{ij}^2] \right] \; .
\eeq
This vacuum state has the property of ``dimensional reduction,'' in the sense that computation of a 
large spacelike loop in $3+1$ dimensions reduces to the calculation of a large Wilson 
loop in $3$ Euclidean dimensions.  Suppose  $\Psi_{0}^{(3)}$  is the ground state of the 3+1 
dimensional theory, and $\Psi_{0}^{(2)}$  is the ground state of the 2+1 dimensional theory.   
If these ground states both have the dimensional reduction form, and $W(C)$ is a large planar 
Wilson loop, then the area law falloff in $D=3+1$
dimensions follows from confinement in two Euclidean dimensions in two steps:
\bea
           W(C) &=& \langle \mbox{Tr}[U(C)]\rangle^{D=4} 
                      = \langle \Psi^{(3)}_{0}|\mbox{Tr}[U(C)]|\Psi_0^{(3)}\rangle
\non \\
                        &\sim&  \langle \mbox{Tr}[U(C)]\rangle^{D=3} 
                      = \langle \Psi^{(2)}_{0}|\mbox{Tr}[U(C)]|\Psi_0^{(2)}\rangle
\non \\
                        &\sim& \langle \mbox{Tr}[U(C)]\rangle^{D=2} 
\eea
In $D=2$ dimensions the Wilson loop can of course be calculated analytically, and we know 
there is an area-law falloff, with Casimir scaling of the string tensions.

   On the other hand, dimensional reduction cannot be exactly right.  For one thing, there is no color screening in 2D Yang-Mills, so we would get the wrong N-ality properties in 2+1 dimensions.  For another, the dimensional reduction state has the wrong high-momentum behavior.  Thus there must be corrections, and in fact it can be shown \cite{Greensite:2007ij}, in the context of strong coupling lattice  gauge theory, that small corrections to the dimensional reduction vacuum wavefunctional are responsible for color screening.  In this talk I will be concerned with the vacuum state in 2+1 dimensions, where the following well-motivated proposals for the ground state have been advanced by various groups:    
\bea
\Psi_{GO}[A] &=& \exp\left[-{1\over 2g^2}\int d^2x d^2y ~ F_{12}^a(x) 
  \left({1 \over \sqrt{-D^2 - \l_0 + m^2}} \right)^{ab}_{xy} F_{12}^b(y) \right]    
\\
     \Psi_{KKN}[A]  &=&  \exp\left[-{1\over 2g^2}\int d^2x d^2y ~ F_{12}^a(x) \left( {1 \over \sqrt{-\nabla^2 + m^2} + m} \right)_{xy} F_{12}^a(y) \right]
\\
\Psi_{hybrid}[A] &=& \exp\left[-{1\over 2g^2}\int d^2x d^2y ~ F_{12}^a(x)   \left( {1 \over \sqrt{-D^2 -\l_0+ m^2} + m} 
\right)^{ab}_{xy} F_{12}^b(y) \right]
\\
 \Psi_{CG} [A] &=& \exp\left[- \frac{1}{2} \int
\frac{d^2k}{(2\pi)^2} \overline{\omega}(k)  A_i^a(k) A_i^a(-k)\right] \; .
\eea
The temporal gauge wavefunctional $\Psi_{GO}$ was suggested by Olejn\'{\i}k and myself \cite{Greensite:2007ij}.  In this equation $-D^2$
is minus the covariant Laplacian, $\l_0$ is its lowest eigenvalue, and $m$ is a parameter with dimensions of mass.  Fixing $m$ to obtain the known string tension, we then find that the mass gap and Coulomb gauge ghost propagator agree with standard Monte Carlo results.  The wavefunctional $\Psi_{KKN}$ was obtained by Karabali, Kim, and Nair (KKN) \cite{Karabali:1998yq} as an bilinear approximation to the ground state within their ``new variables'' approach.  When this approximation is transformed back to the usual variables in temporal gauge, the result is the expression shown above.  In their case $m = g^2 N / 2 \pi$.  If one simply drops the term $-\nabla^2$, then the string tension agrees remarkably well (in the large $N$ limit) with the Monte Carlo result.  A questionable feature of this procedure is that $\Psi_{KKN}[A]$ is not gauge-invariant as it stands, and is therefore not a physical 
state.   A reliable computation of the string tension via a dimensional reduction approximation really requires, and depends on the choice of, a gauge-invariant completion of this state, which goes beyond the bilinear result offered by 
KKN.\footnote{For example, a naive replacement of $-\nabla^2$ by $-D^2$ in $\Psi_{KKN}$ would lead to an infinite string tension in the continuum limit, cf.\  \cite{Greensite:2007ij}.} A gauge-invariant ground state combining features of $\Psi_{GO}$ and $\Psi_{KKN}$ is the hybrid state $\Psi_{hybrid}$, suggested in \cite{Greensite:2011pj}, which agrees with $\Psi_{KKN}$ when evaluated on abelian ($[A_1,A_2]=0$) configurations.  Both $\Psi_{GO}$ and $\Psi_{hybrid}$ have the dimensional reduction property, when evaluated on long-wavelength configurations.
The Coulomb-gauge wavefunctional $\Psi_{CG}$ has been advocated and developed by Szczepaniak and co-workers and by Reinhardt and co-workers, cf.\ eg.\  \cite{Szczepaniak:2001rg}
and \cite{Feuchter:2004mk}. In their proposal $\overline \omega(k)$ is determined by a set of coupled integral equations.  This approach leads to an enhancement of the Coulomb ghost propagator, and a confining Coulomb potential.  On the other hand, an area law for spacelike Wilson loops is not obtained.

   All of these proposals have the same free-field behavior for field configurations with only short-wavelength, quasi-abelian ($[A_1,A_2] \approx 0$) components.  They differ at the long-wavelength end.  Therefore we would like to compare the predictions of these proposed wavefunctionals with those of the true vacuum wavefunctional $\Psi_{true}[A]$ evaluated on configurations which are either very long wavelength abelian, or essentially non-abelian.\footnote{It should be noted that the true vacuum wavefunctional in Coulomb gauge is the same as the true ground state in temporal gauge, when both are evaluated on field configurations fixed to Coulomb gauge.}

\section{The Measurement Method}

     Consider a modified lattice Monte Carlo procedure in $D$ Euclidean dimensions, in which the lattice fields on a particular time slice, $t=0$ say, are restricted to be one of a finite set of $D-1$ dimensional lattice configurations ${\cal U} \equiv \{U^{(m)}_k(\bx),m=1,2,...,M\}$.  Link variables at time $t\ne 0$ are updated normally, but on the time slice $t=0$ a member of the set  ${\cal U}$ is selected at random, and accepted or rejected according to the Metropolis algorithm.
Let $N_n$ be the number of times that the $n$-th member of the set is accepted.  Then it is not hard to show that 
\cite{Greensite:2011pj}
\beq
           {\Psi^2_0[U^{(n)}] \over \Psi^2_0[U^{(m)}]} = \lim_{N_{tot} \ra \infty} {N_n \over N_m} \; ,
\eeq
where $\Psi_0[U]$ is the ground state of the lattice transfer matrix in temporal gauge.  Thus we are able, in principle, to compute the relative amplitudes of the \emph{true} Yang-Mills vacuum wavefunctional, in any given set of configurations.\footnote{This method was first suggested in ref.\ \cite{Greensite:1987rg}.}  This means that,
given any proposal
 \beq
          \Psi_{proposal}[U] = \N e^{-\oh R[U]} 
\eeq
for the Yang-Mills vacuum, all we have to do is to plot
\beq
   -\log\left[{N_m \over N_{tot}}\right]   ~~ \mbox{vs.} ~~ R[U^{(m)}] \; ,
\eeq     
where $N_{tot}$ is the total number of updates at $t=0$, and serves as a convenient normalization.  If the proposal is correct, then the data should fall on a straight line with a slope $=1$.

\section{Results}
  
     We will consider three types of trial configurations.  The first is abelian plane waves
\bea
            U^{(m)}_1(n_1,n_2) &=& \sqrt{1 - (a^{(m)}(n_2))^2} \mathbbm{1}_2 + i a^{(m)}(n_2)
            \s_3 \;
\non \\
            U^{(m)}_2(n_1,n_2) &=&    \mathbbm{1}_2
\non \\
             a^{(m)}(n_2) &=& {1 \over L} \sqrt{\a + \g m} \,\cos\left({2 \pi n_2 \over L}  \right) \; ,
\label{planewavesdef}
\eea
where  $m=1,2,...,m_{max}$  with $L$ the lattice extension, $\a,\g$ some constants, and $\tk^2 = 2\Bigl(1-\cos\left({2\pi \over L}\right)\Bigr)$ is the squared lattice momentum.  To convert to physical units, we will arbitrarily set
the string tension to $\s=(440~\mbox{MeV})^2$, and then the lattice spacing is $a=\sqrt{\s_{lat}/\s}$ as usual.  For the
GO, KKN, and CG wavefunctionals one has
\beq
 -\log\left({N_n \over N_{tot}}\right) = 2 (\a +\g n) \omega(\tk^2) + r_0 \; ,
\label{fit}
\eeq
with the predictions
\beq
           \omega(k^2) =   \left\{ \begin{array}{cl}
           {1\over g^2}  {k^2 \over \sqrt{k^2 + m^2}} & \mbox{GO} \cr
  {1\over g^2}  {k^2 \over \sqrt{k^2 + m^2} + m} & \mbox{KKN} \end{array} \right. \; .
\eeq
For CG, a numerical approach is required.  In that case the integral equations contain some constant parameters
$c_1,c_2$, and we have $\omega(0)=0$ for $c_1=0$, with $\omega(0)>0$ otherwise.

\begin{figure}[t!]
\centerline{\scalebox{0.72}{\includegraphics{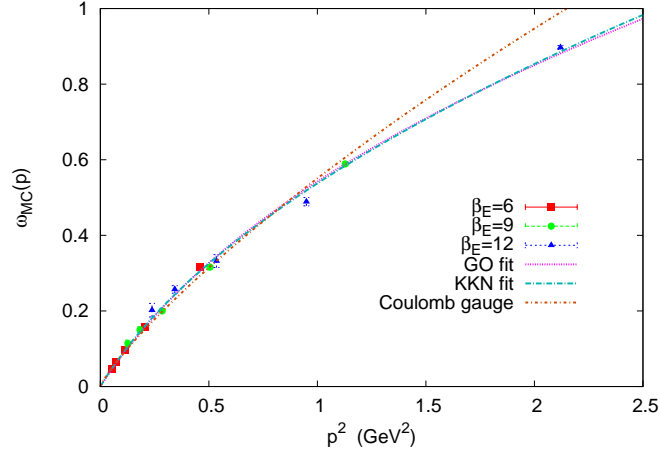}}}
\caption{Cumulative data for $\omega_{MC}$ vs.\ $p^2$ in physical units, compared to theoretical predictions.}
\label{om}
\end{figure}

   From a best linear fit to the lhs of \rf{fit} at a given value of $\tk^2$, we can extract $\omega(\tk)$ from the data, and
compare these values, obtained at various $\tk^2$, to the theoretical predictions.  For GO and KKN there is a dimensionless parameter $g^2/m$ which can be chosen from a best fit to the data.  Our results are shown in Fig.\ \ref{om} for lattices of extensions  $L=16,24,32,40,48$, and Euclidean lattice couplings $\b_E=6,9,12$ for the numerical simulations. The GO and KKN (= hybrid for abelian) proposals both work well, and are indistinguishable in this range of momenta.  The CG proposal also works well, for the choice (shown here) of  $c_1=0$.

\begin{figure}[htb]
\centerline{\scalebox{0.65}{\includegraphics{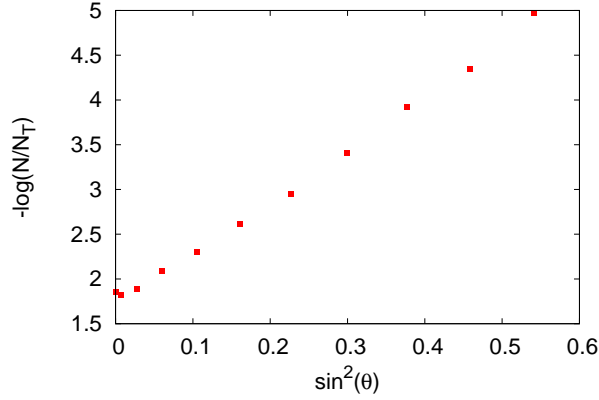}}}
\caption{Dependence of $-\log(N_n/N_T)$ on the "non-abelianicity" of the non-abelian constant configurations, determined by $\sin(\theta_n)$.}
\label{fig1}
\end{figure}

    Next we consider non-abelian constant configurations, beginning with configurations in which links $U_1$ and
$U_2$ have a fixed amplitude, but whose commutator $[U_1,U_2]$ depends on an angle $\theta$, which varies in 
the set:
\bea
      U^{(m)}_1(n_1,n_2) &=& \sqrt{1-\a^2}\mathbbm{1}_2  + i\a \s_1
\non \\
      U^{(m)}_2(n_1,n_2) &=& \sqrt{1-\a^2}\mathbbm{1}_2 + i\a(\cos(\theta_m) \s_1 + \sin(\theta_m) \s_2) \; ,
\label{nac2}
\eea 
and we take evenly spaced $\theta_m = \g (m-1)$.  For the GO and hybrid wavefunctionals, the exponent is proportional to the field-strength squared, i.e. $R[U] \propto (A_1 \times A_2)^2$, for small $\a$, and therefore
\beq
            R_{GO,hybrid}[U^{(n)}] \propto \sin^2(\th_n)
\eeq
for these field configurations.
For the Coulomb gauge  wavefunctional, however, the exponent is proportional to the gauge-field squared, i.e. $R[U] \propto A_1^2 + A_2^2$, and hence, since the amplitudes of
$A_1$ and $A_2$ are fixed in the set \rf{nac2},
\beq
             R_{CG}[U^{(n)}] \propto \overline{\omega}(0) \; ,
\label{RCG0}
\eeq
independent of the angle $\th_n$.    If $\overline{\omega}(0)=0$, which seems optimal for agreement with the
plane wave data, then $R_{CG}$ would also be independent of the amplitude of the gauge fields.  These predictions are easy to check, since $-\log(N_n/N_{tot})$ should be equal to $R[U]$ plus a constant.  Plotting $-\log(N_n/N_{tot})$ vs.\
$\sin^2(\th_n)$, we find the result shown in Fig.\ \ref{fig1}.  This result is consistent with the GO and hybrid proposals,
but is clearly not compatible with CG.  

    Finally, we consider sets of non-abelian constant configurations with maximal non-abelianicity, i.e.\ $\th=\pi/2$,
but varying amplitude:
\bea
            U^{(m)}_1(n_1,n_2) &=& \sqrt{1 - (a^{(m)})^2} \mathbbm{1}_2 + i a^{(m)} \s_1
\non \\
            U^{(m)}_2(n_1,n_2) &=&  \sqrt{1 - (a^{(m)})^2} \mathbbm{1}_2 + i a^{(m)} \s_2
\non \\
              a^{(m)} &=& \left[ {\a + \g m \over 20 L^2} \right]^{1/4} \; .
\label{nac1}
\eea
We then plot $\log N_m/N_{tot}$ vs.\ $R[U^{(m)}]$, to see whether the data points fall on a straight line and, if so, whether the slope equals one. 
 
\begin{figure}[t!]
\begin{center}
\subfigure[]  
{   
 \label{nac-a}
 \includegraphics[scale=0.53]{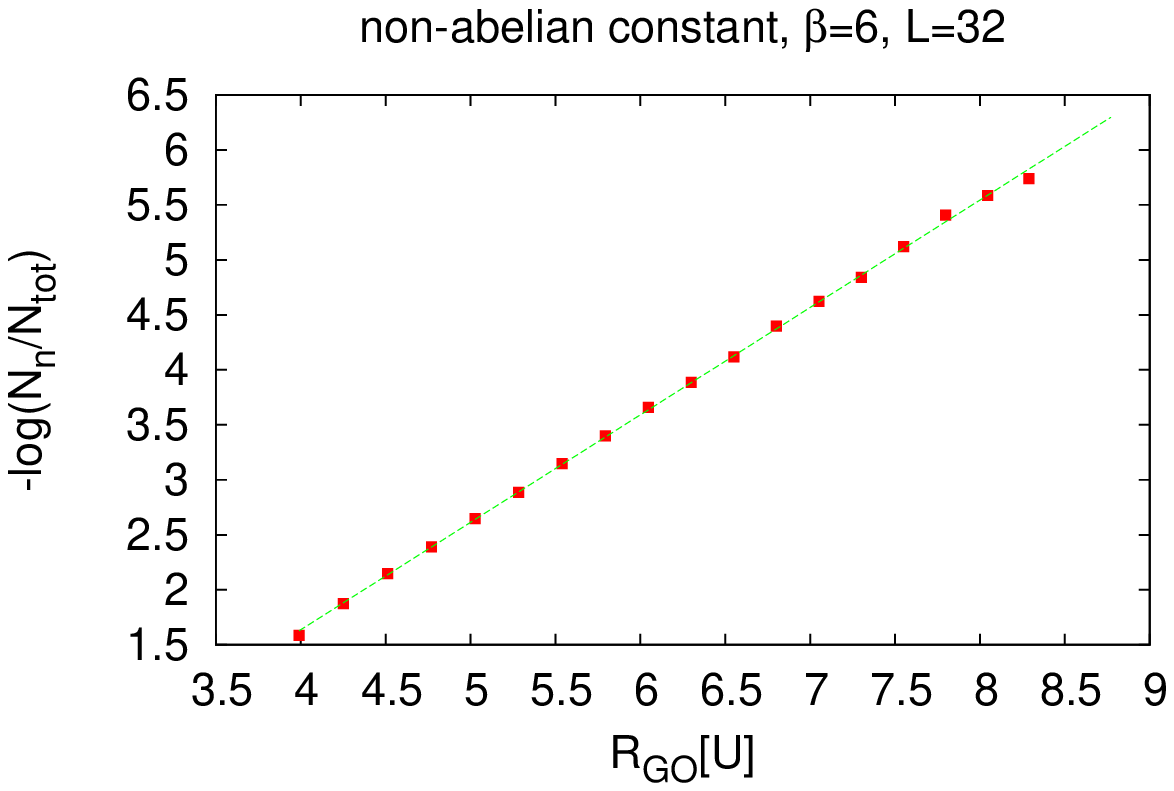}
}
\hspace{0.5cm}
\subfigure[]   
{  
 \label{nac-b}
 \includegraphics[scale=0.53]{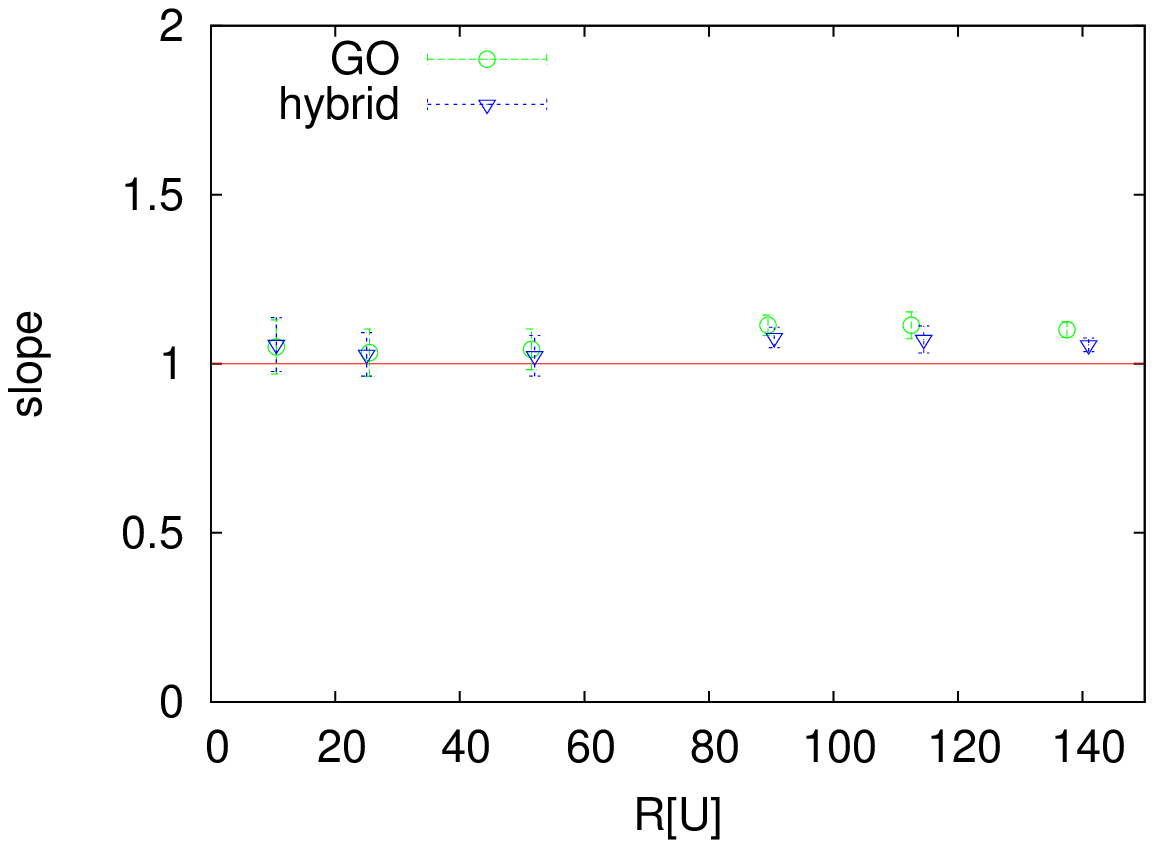}
 }
\end{center}
\caption{ (a) Plot of $-\log(N_n/N_T)$ vs.\ $R_{GO}$ for non-abelian constant configurations, maximal non-abelianicity,
at $\b_E=6,~L=32,~\a=2,~\g=0.15$.  In this case the straight line fit has a slope = 0.98.  (b) $\b_E$=12 calculation, for both types of wavefunctionals.}
\label{nac} 
\end{figure}

   An example of the $-\log[N_n/N_{tot}]$ vs.\ $R_{GO}[U^{(n)}]$ data at $\b_E=6$ is shown in Fig.\ \ref{nac-a}, for the
choice $\a=2, \g=0.15$.  Although the data is nicely fit by a straight line which has a slope close to unity, this fact must be interpreted with caution because, since the number $N_n$ falls off exponentially with $R_{GO}[U^{(n)}]$, the range of $R$ must necessarily be kept small; typically $\D R \approx 4-5$.  This {\it  could} mean that the tendency of the data to lie on a straight line is misleading, and perhaps we are simply looking at the tangent of a curve.  It is therefore necessary to extract the slope of the straight line over small intervals centered around points over a wide range of $R$.  The slope of the data vs.\ $R[U]$ is displayed in Fig.\ \ref{nac-b} for both the GO and hybrid proposals, and we see that in both cases the slope is close to
unity.

\section{Numerical Simulation of the GO and hybrid wavefunctionals}

    It is also possible to carry out numerical simulations with the probability distribution $\Psi_0^2[U]$ for the GO and hybrid proposals, via a method developed in ref.\ \cite{Greensite:2007ij}.  Using this method, one can compute the 
mass gap, extracted from the gauge-invariant $\langle F_{12}^{a2}(x) F_{12}^{a2}(y) \rangle$ correlator, as well as gauge-dependent quantities such as the Coulomb gauge ghost propagator 
\beq
          G(R) =  - \left\langle {1 \over \nabla \cdot D} \right\rangle^{aa}_{|x-y|=R} \; ,
\eeq
and the color Coulomb potential
\beq
V_C(R) = - \left\langle {1\over \nabla \cdot D} (-\nabla^2) {1\over \nabla \cdot D} \right\rangle^{aa}_{|x-y|=R} \; .
\label{VC}
\eeq

\begin{figure}[t!]
\begin{center}
\subfigure[]  
{   
 \label{gap}
 \includegraphics[scale=0.53]{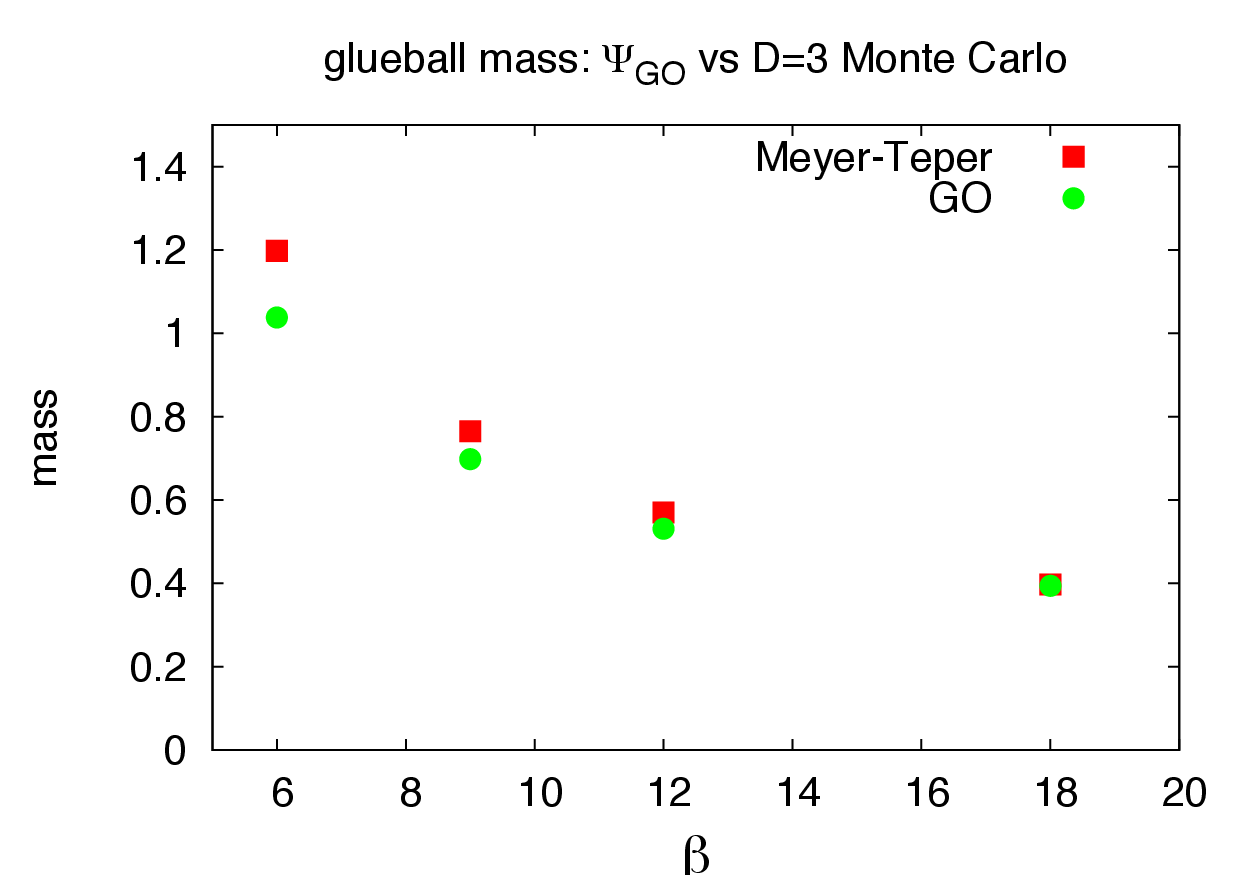}
}
\hspace{0.5cm}
\subfigure[]   
{  
 \label{ghost}
 \includegraphics[scale=0.53]{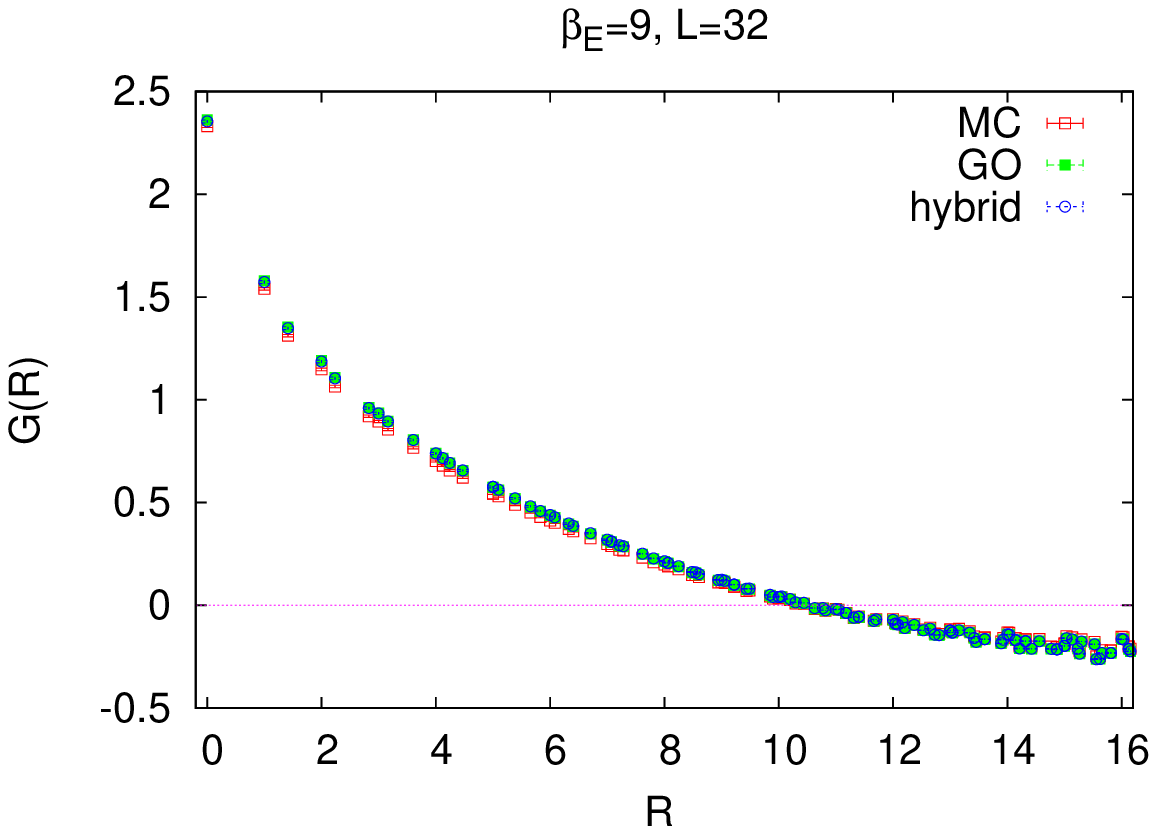}
 }
\end{center}
\caption{ (a) Mass gap extracted from simulations of the GO wavefunctional, compared to the results obtained by Meyer and Teper \cite{Meyer:2004jc}. (b)The ghost propagator derived from standard Monte Carlo (MC) simulation at $\b_E=9$, and
the same quantity calculated by simulation of the GO and hybrid wavefunctionals.}
\label{cg} 
\end{figure}

\begin{figure*}
  \begin{center}
    \begin{tabular}{cc}
      \resizebox{50mm}{!}{\includegraphics{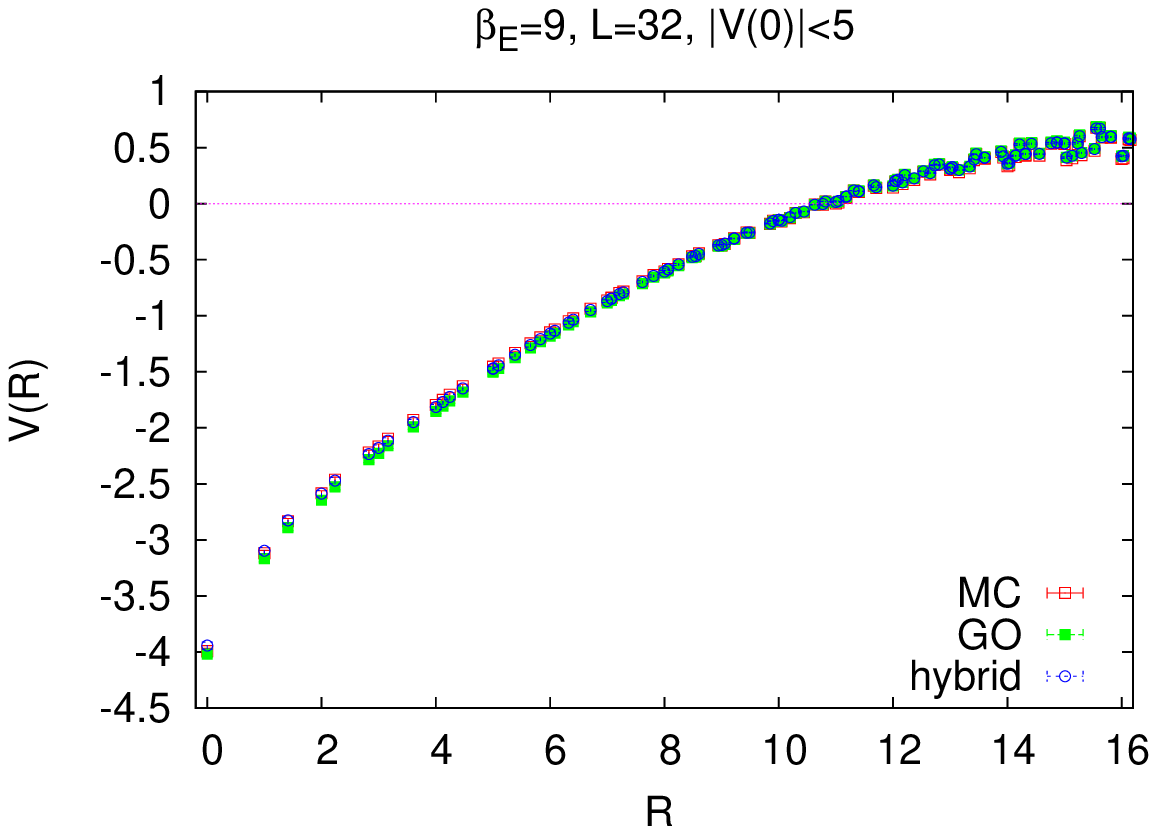}} &
      \resizebox{50mm}{!}{\includegraphics{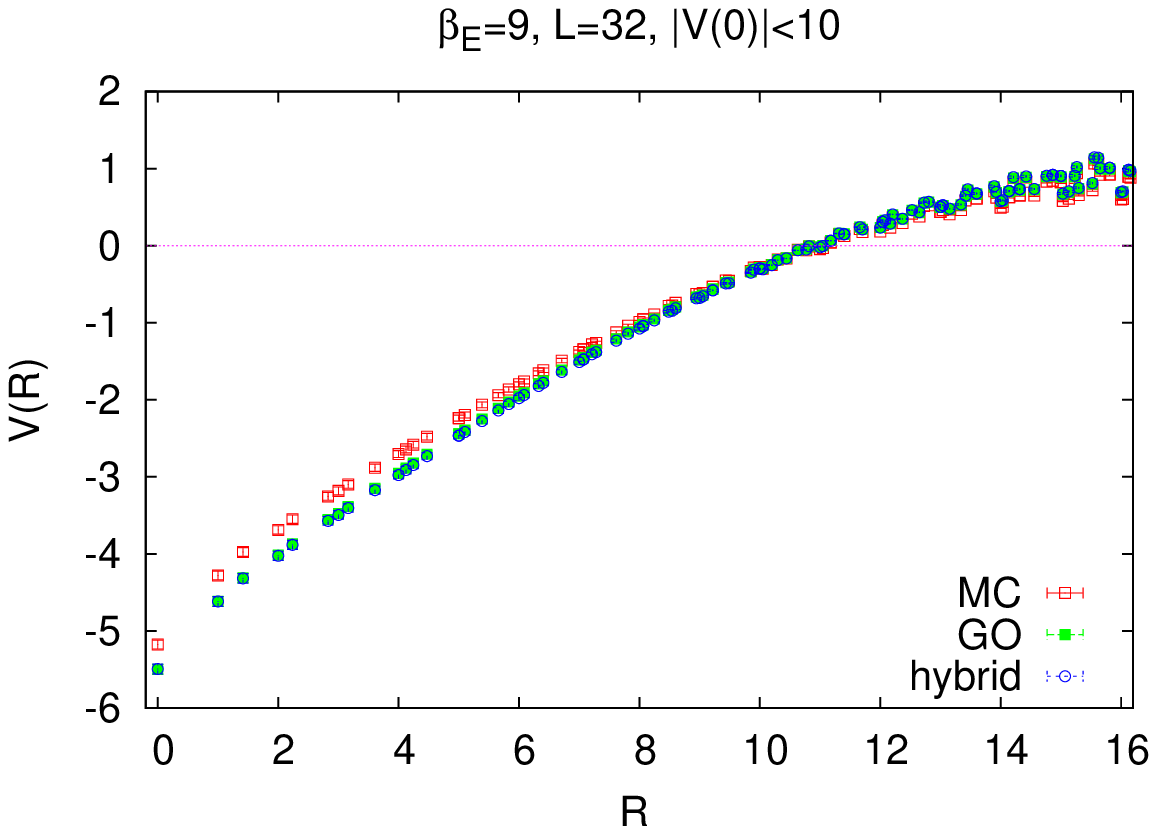}} \\
      \resizebox{50mm}{!}{\includegraphics{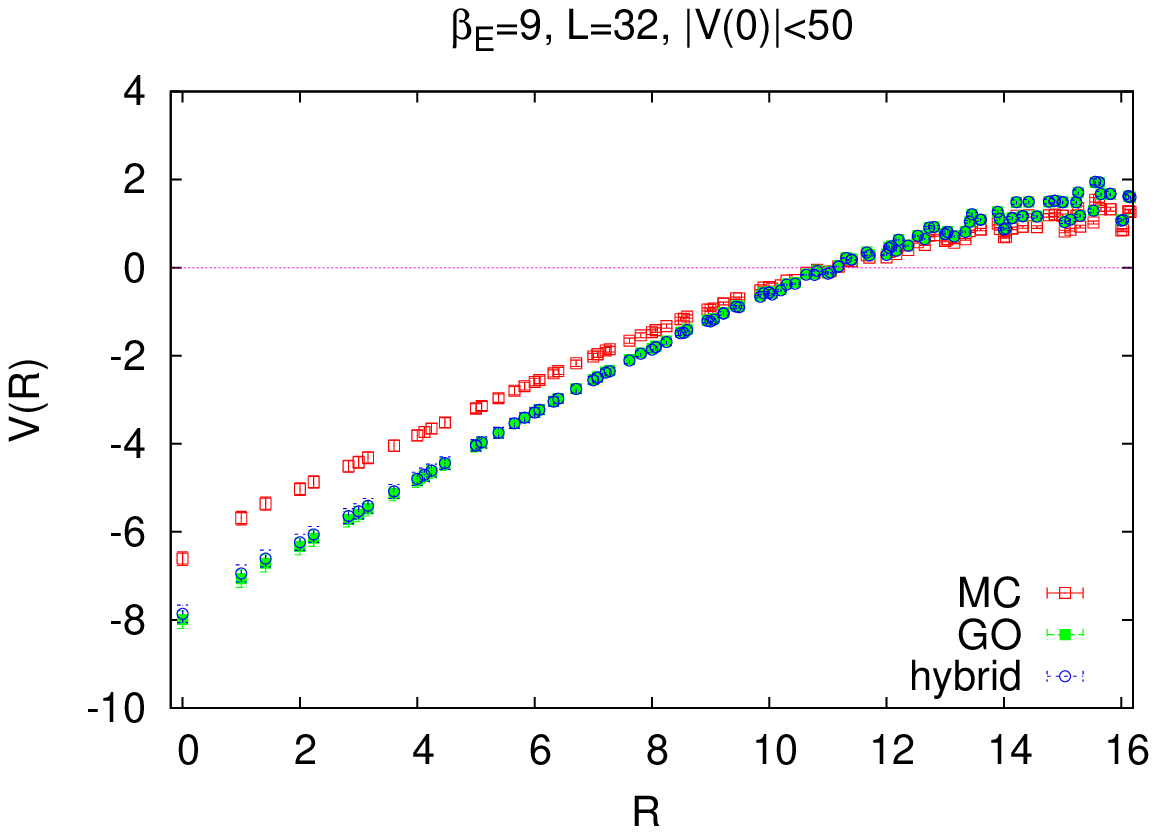}} &
      \resizebox{50mm}{!}{\includegraphics{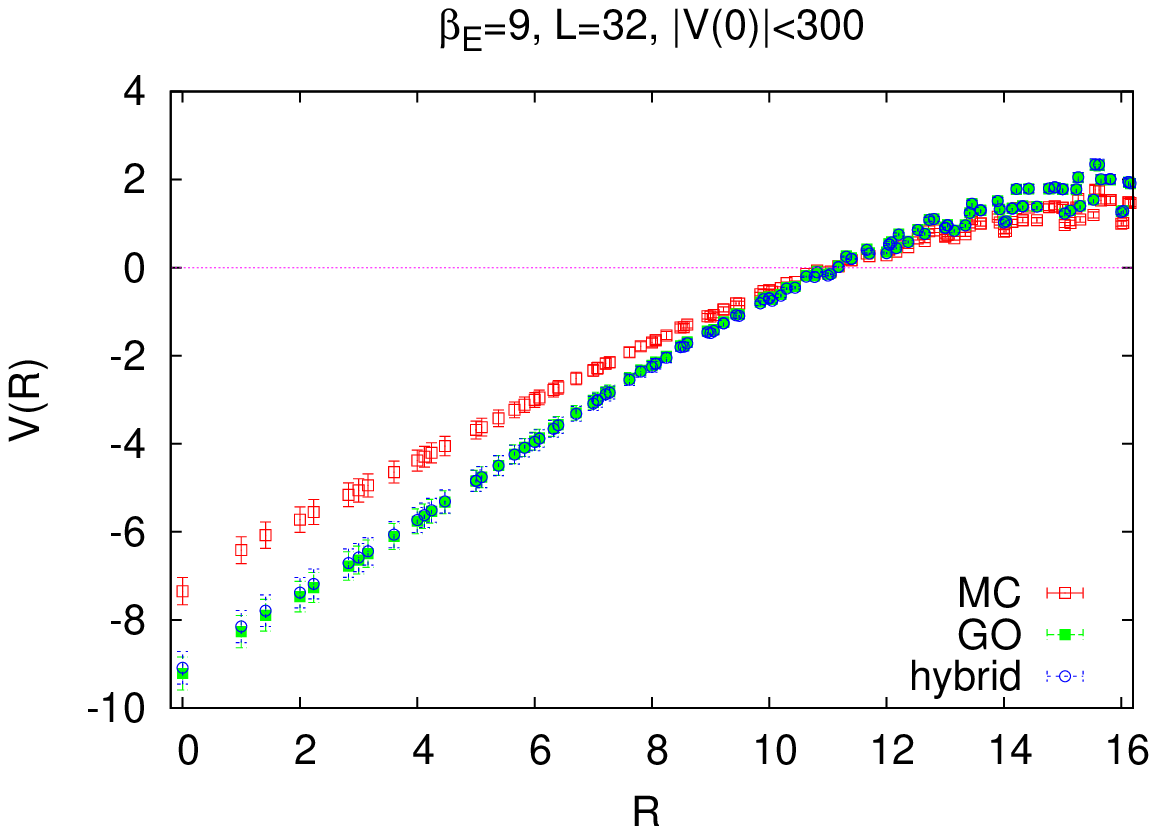}} \\
    \end{tabular}
    \caption{Data for the Coulomb potential at $\b_E=9$ and $L=32$, derived from
    MC, GO and hybrid simulations, with a cut on the data, discarding configurations for
    which $|V(0)|$ is greater than 5, 10, 50, and 300, respectively.}
    \label{Vc}
  \end{center}
\end{figure*}

    In the case of the mass gap, the values found from $\Psi_{GO}$ appear to converge nicely, in the continuum
limit, to the values found by standard Monte Carlo techniques in ref.\ \cite{Meyer:2004jc}, as seen in Fig.\ \ref{gap}.  An even more striking agreement is found between the Coulomb gauge ghost propagator, extracted from the probability distributions of $\Psi^2_{GO}$ and $\Psi^2_{hybrid}$, and the same quantity computed by the usual Monte Carlo simulation techniques in three Euclidean dimensions.  These are displayed in Fig.\ \ref{ghost}.   Agreement for the 
color Coulomb potential is not as good, but we have traced the disagreement to the fact that this quantity is extremely
sensitive to infrequent ``exceptional" configurations, in which the value of $\l_0$ is unusually low.  These rare cases can be eliminated by imposing a lower bound cutoff on the value of $V_C(0)$ obtained from a single configuration.  As the lower bound is raised, agreement between the GO/hybrid and Monte Carlo values is restored, as seen in Fig.\ \ref{Vc}.  The color Coulomb potential 
involves two powers of the inverse Faddeev-Popov operator, while the ghost propagator is a single inverse power.  We interpret our results to mean that $\Psi^2_{GO}$ and $\Psi^2_{hybrid}$ are close to the true ground state $\Psi^2_{true}$ for the bulk of the probability distribution, but may deviate somewhat in the tail of the distribution. 

\section{Is the Horizon enough?}

    I will now turn to the second topic mentioned in the Introduction, relating to the Gribov horizon.  We have seen that
the GO and hybrid wavefunctionals provide an area-law falloff for spacelike Wilson loops, as well as a confining Coulomb potential, and pass the tests described in previous sections.  But do we really need such sophisticated wavefunctionals to obtain a confining Coulomb potential?   Is it not possible that a simple gaussian distribution for momentum-space field components is sufficient for Coulomb confinement, provided the gauge fields are close to the Gribov horizon?  

  We recall that the Gribov horizon is a cutoff in the functional integral over transverse A-fields, which is simply ignored in ordinary perturbation theory.  The domain of the functional integral should be restricted to configurations for which the Faddeev-Popov operator ${M = -\nabla \cdot D}$ is positive definite, i.e.\  the lowest non-trivial eigenvalue $\l_0$ of $M$
is positive.  This domain is known as the ``Gribov region,'' and it is bounded by first ``Gribov horizon,'' where $\l_0=0$. 
In the Gribov-Zwanziger picture of confinement, most of the volume of the Gribov region is concentrated very near the horizon, and therefore $\l_0$ is typically close to zero.  Because the Coulomb potential 
$\langle M^{-1}(-\nabla^2)M^{-1}\rangle$
contains two inverse powers of the Faddeev-Popov operator, this proximity of typical gauge field configurations to the horizon is conjectured to enhance the Coulomb potential from $1/r$ to linear behavior.  It was also argued by Gribov  that the
restriction to the Gribov region would suppress the transverse gluon propagator in the infrared, such that in Coulomb
gauge, at equal times,
\beq
           D^{ab}_{ij}(k) =  {\d_{ij} - {k_i k_j \over k^2} \over 2 \sqrt{ k^2 + m^4/k^2}} \d^{ab} \; ,
\eeq
where $m$ is a constant with dimensions of mass, and $k=|\vec{k}|$ refers to the spatial components of momentum.  

This leads to an interesting question:  Suppose we generate transverse gauge fields which are (arbitrarily) close to the horizon, and which result in whatever transverse propagator (e.g. Gribov's) is desired.  Would such gauge fields result in a confining Coulomb potential?   To answer this question, let us generate some large set of random numbers from a normal distribution, and use these random numbers to construct transverse momentum-space gauge fields $A_i(k)$  with the required properties.  Then we calculate the Coulomb potential, and investigate how that potential
depends on the proximity to the Gribov horizon, and the form of gluon propagator chosen.\footnote{Related work has
been carried out in refs.\ \cite{Matevosyan:2008nf} and \cite{Quandt:2011zz}.  In those articles the $A$-fields are drawn from a probability distribution corresponding to a Gaussian wavefunctional in 3+1 dimensions, leading to a particular Coulomb gauge gluon propagator (infrared finite in the former case, Gribov form in the latter).   Transverse gauge fields drawn from these distributions were used to calculate the Coulomb gauge ghost propagator numerically, and the results showed at best a modest infrared enhancement of the ghost dressing function.  Our numerical approach is similar, except that we adjust the proximity to the horizon, and calculate the color Coulomb potential (in 2+1 dimensions) rather than the ghost propagator.}

   In order to calculate the Coulomb potential, we only need the gauge field at a fixed time.  In what follows I will use continuum notation, but lattice regularization, and lattice momenta, are implicit.   From transversality in two dimensions we may write
\beq
         A^a_j(k) = \epsilon_j(k) A^a(k)  \; ,
\eeq
where $\epsilon_j(k)$ is the polarization vector, and we want to select $A^a(k)$ stochastically from a probability distribution such that, on an $L \times L$ lattice    
\beq
\langle A^a_i(k) A^b_j(k') \rangle =  \d^{ab} { \d_{ij} - {k_i k_j \over k^2} \over  \o(k)} L^2 \d_{k,-k'} \; ,
\eeq
where
\beq
 \o(k) = \left\{ \begin{array}{cl}
       2 \sqrt{k^2 + m^4/k^2} & \text{Gribov propagator} \\ \\
       2 \sqrt{k^2 + m^2} & \text{massive propagator} \end{array} \right. \; .
\eeq
This is achieved by choosing, for each momentum and color component,
\beq
 A^a(k) = {L \over \sqrt{2\o(k)}}[\eta^a_1(k) + i\eta^a_2(k)] \; ,
\eeq
where $\eta^a_{1,2}$ are random numbers taken from a normal distribution.  We then set
$A^a_j(k) = \epsilon_j(k) A^a(k)$,  Fourier transform back to position space, and calculate the
Faddeev-Popov matrix ${(-\nabla \cdot D)^{ab}_{xy}}$. 

    The mass parameter  $m$  in  $\o(k)$  can be adjusted to make the lowest (non-trivial) eigenvalue  $\lambda_0$
of the Faddeev-Popov operator positive, and as close to zero as desired.  Then, using the resulting gauge field,  we calculate the Coulomb potential \rf{VC}
at $\beta=4$, averaging over many such configurations.   The question is whether this potential is confining for, e.g., a
distribution which produces a tranverse gluon propagator as proposed by Gribov, given a sufficiently small
value for $\l_0$.

\begin{figure}[t!]
\begin{center}
\subfigure[]  
{   
 \label{gribov-a}
 \includegraphics[scale=0.45]{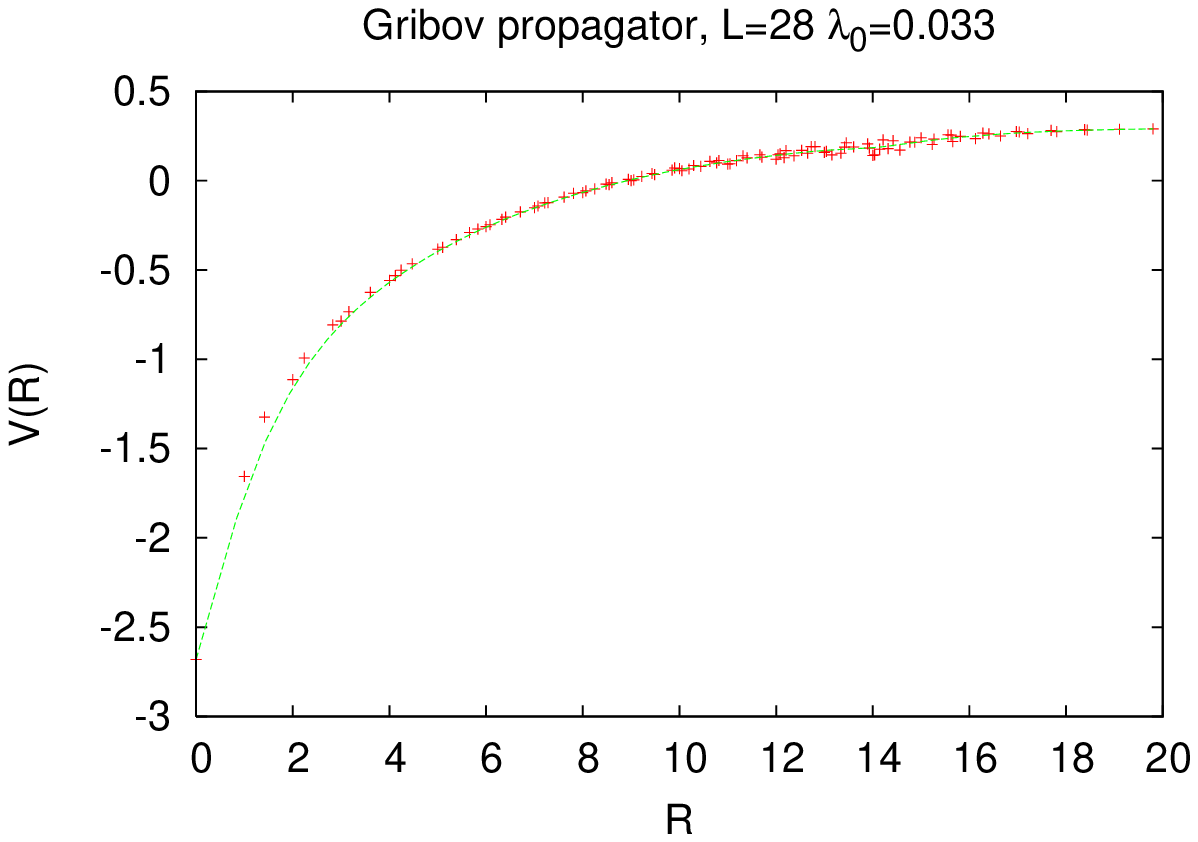}
}
\subfigure[]   
{  
 \label{gribov-b}
 \includegraphics[scale=0.45]{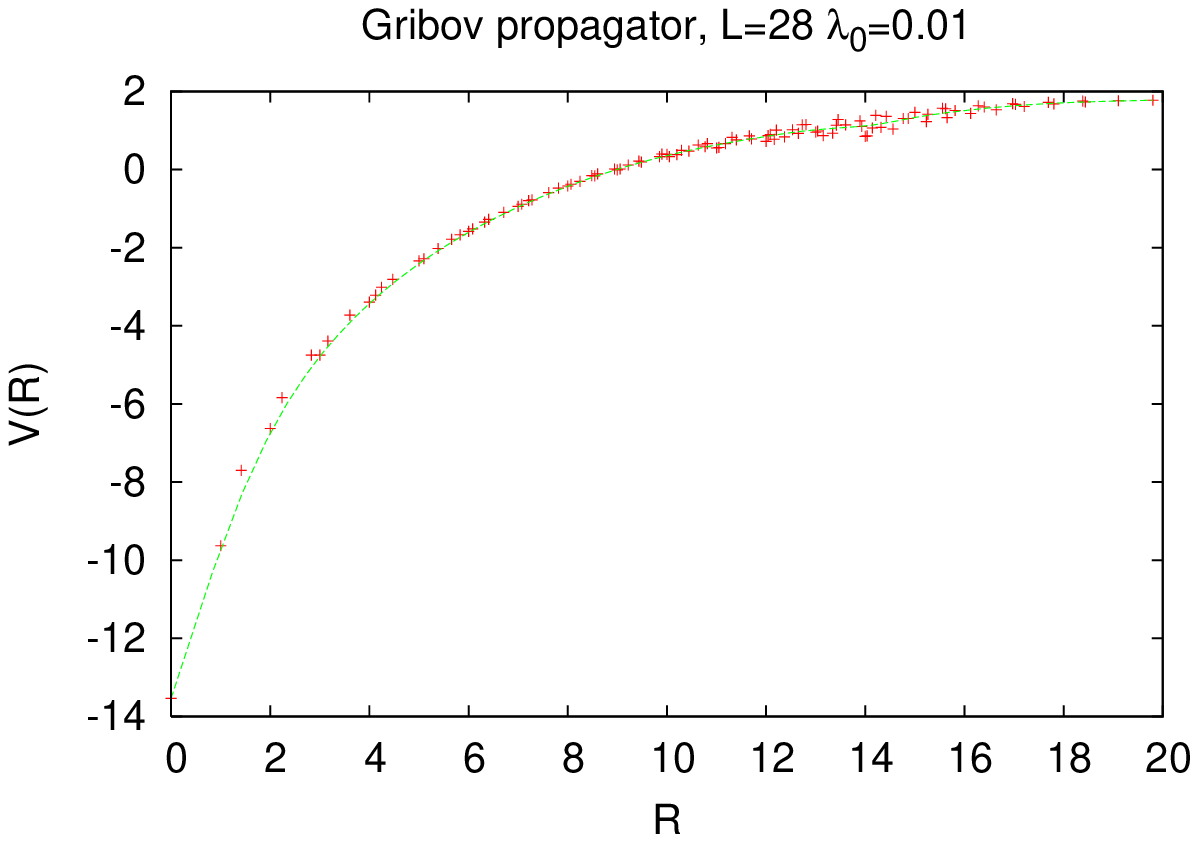}
 }
\subfigure[]   
{  
 \label{gribov-c}
 \includegraphics[scale=0.45]{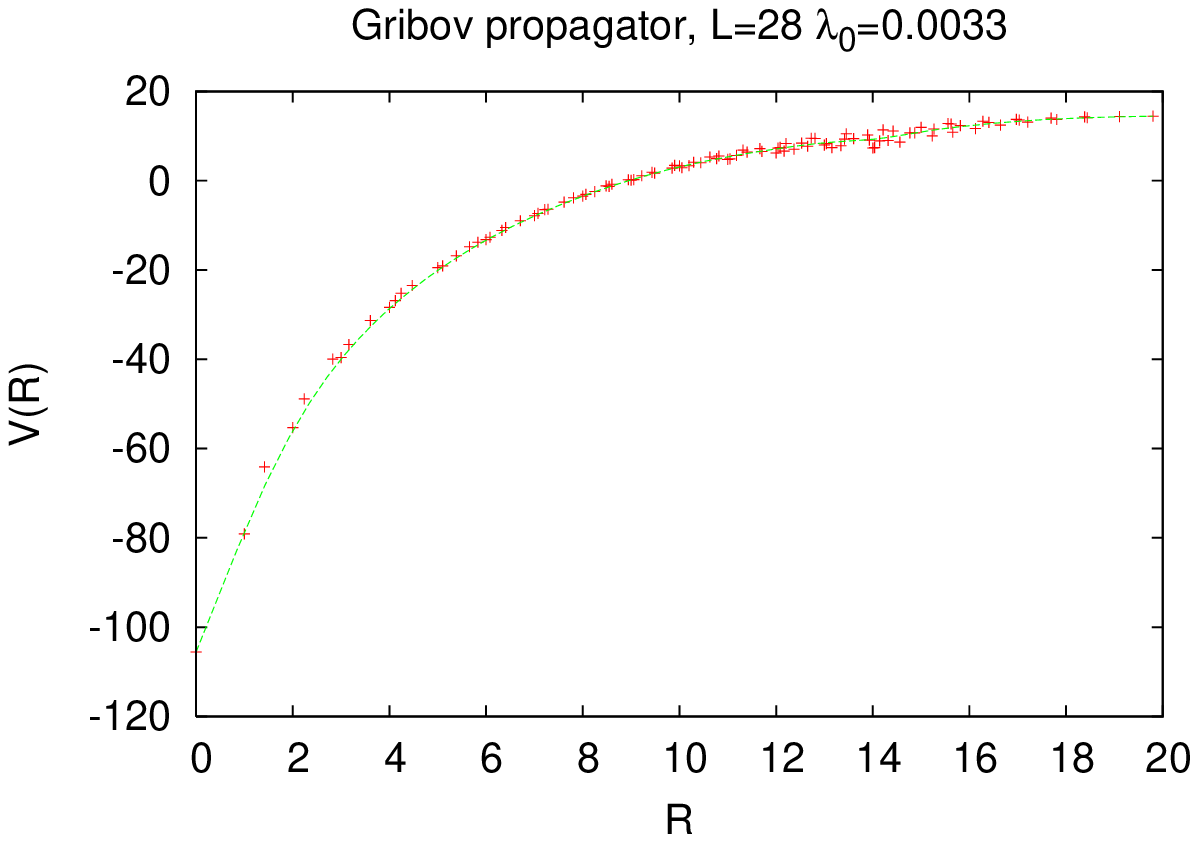}
 }
\subfigure[]   
{  
 \label{gribov-d}
 \includegraphics[scale=0.45]{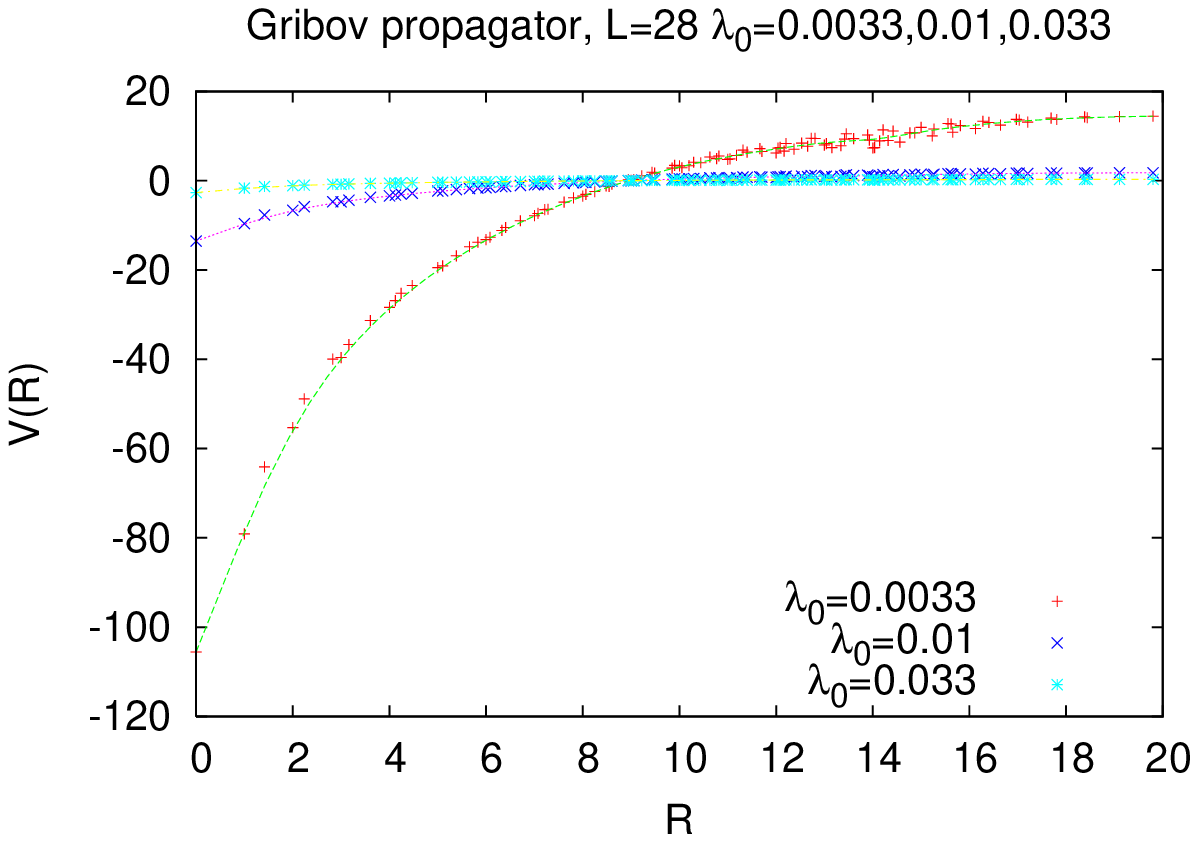}
 }
\end{center}
\caption{Coulomb potential from the Gribov transverse gluon propagator with $\l_0$ = (a) 0.033, (b) 0.01, (c) 0.0033,
(d) combined.}
\label{gribov} 
\end{figure}

\begin{figure}[h!]
\begin{center}
\subfigure[]  
{   
 \label{mass-a}
 \includegraphics[scale=0.45]{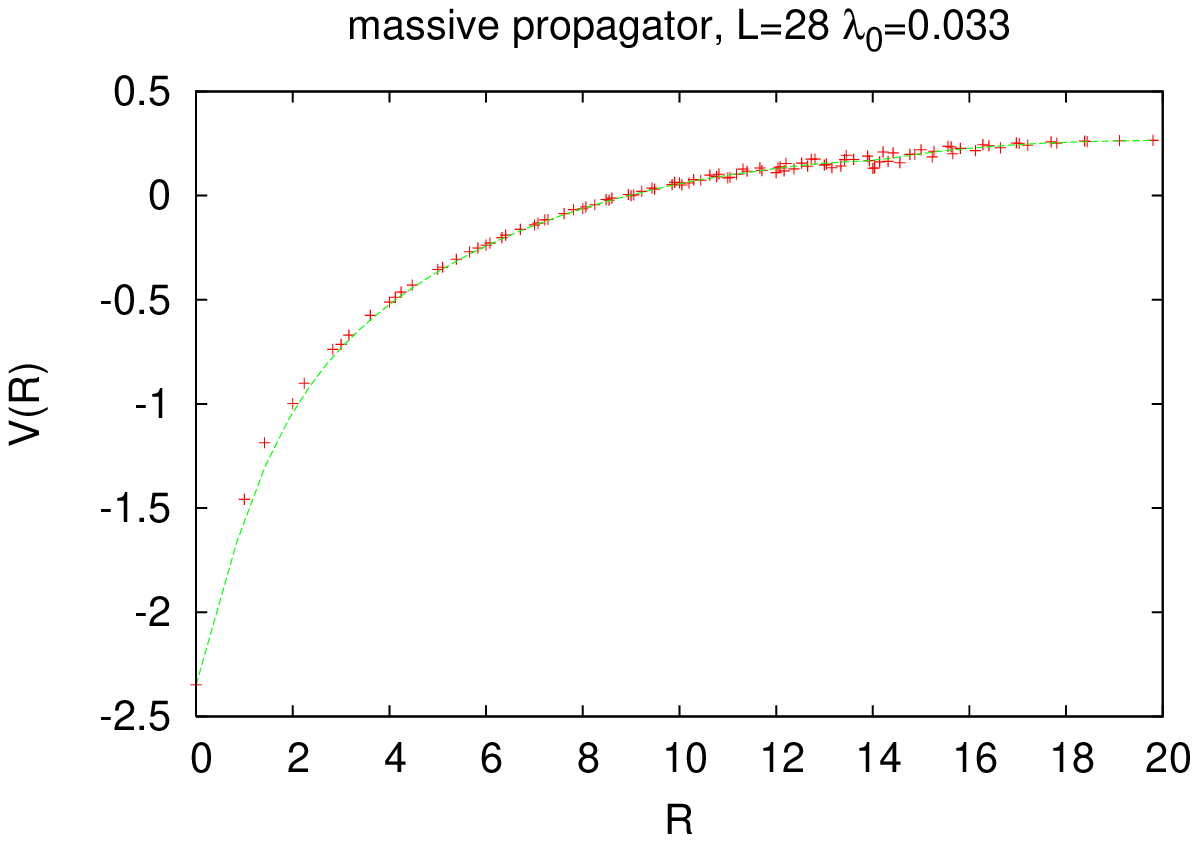}
}
\subfigure[]   
{  
 \label{mass-b}
 \includegraphics[scale=0.45]{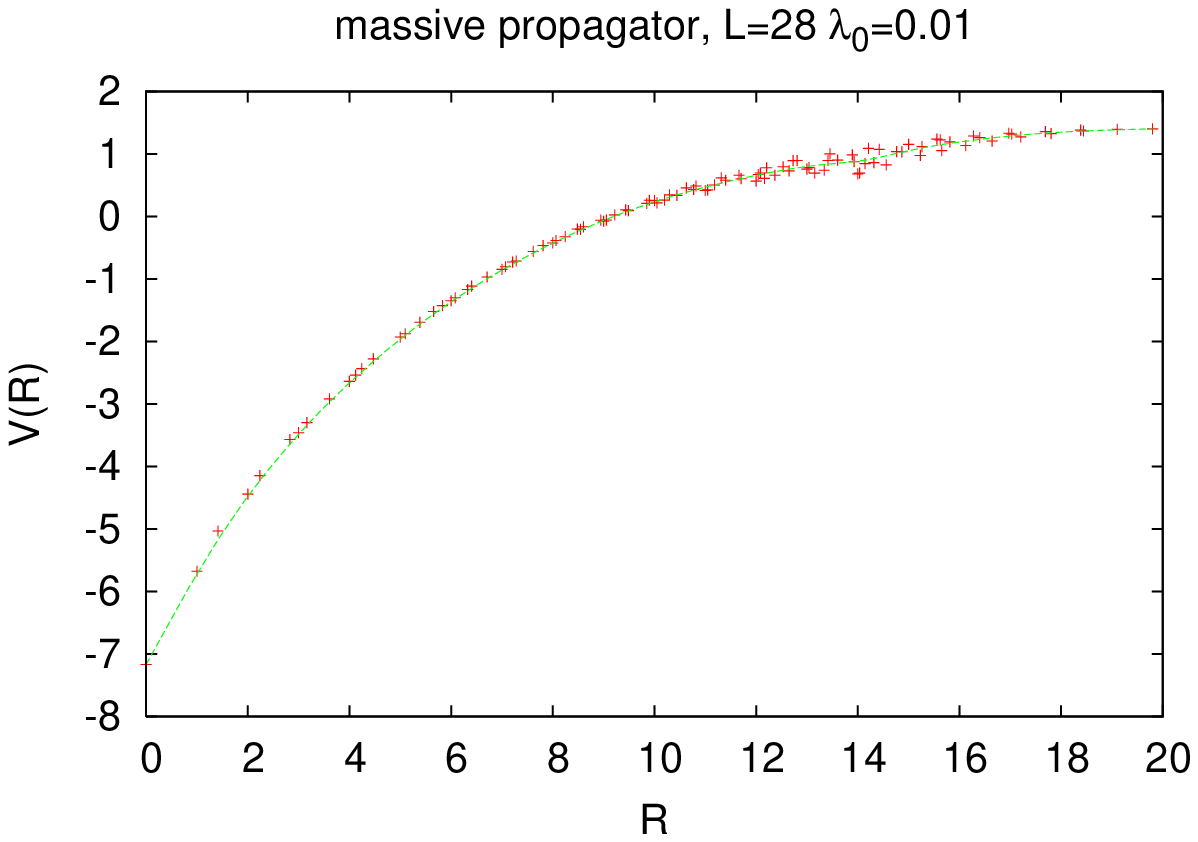}
 }
\subfigure[]   
{  
 \label{mass-c}
 \includegraphics[scale=0.45]{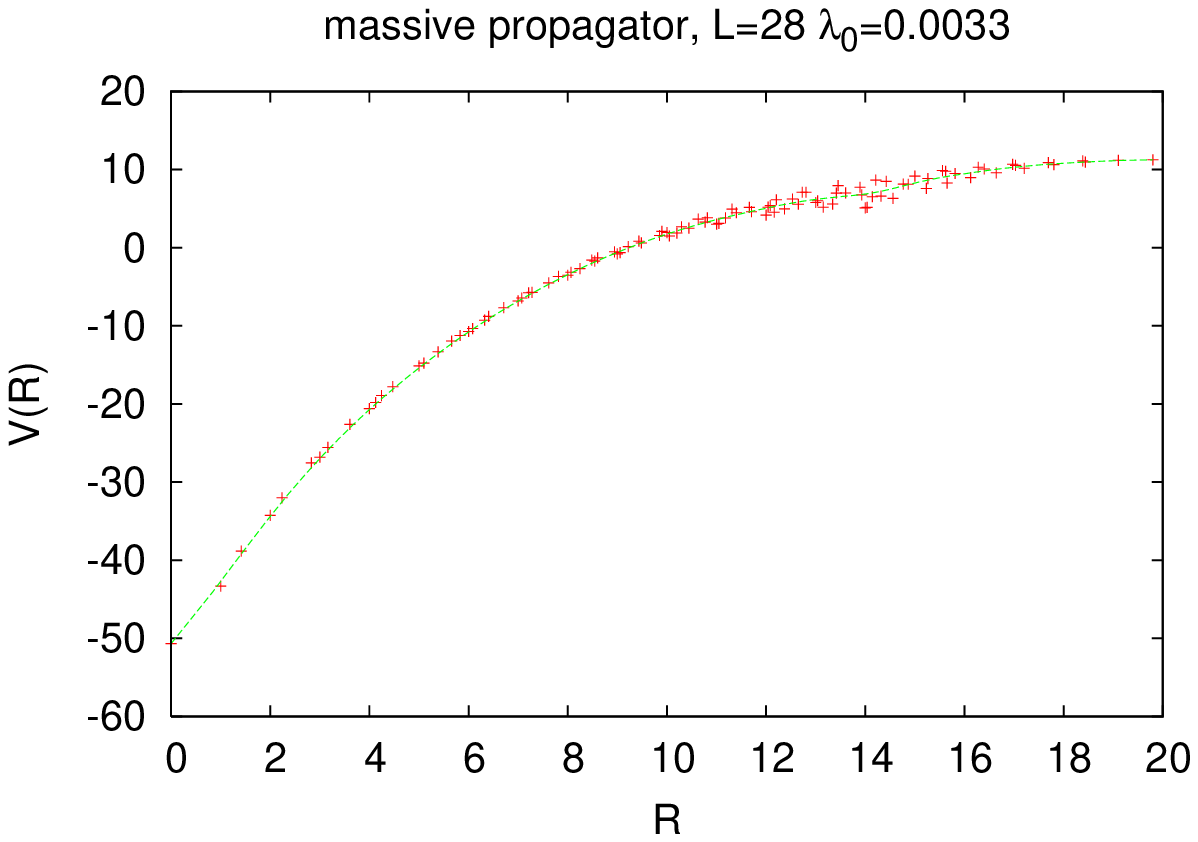}
 }
 \subfigure[]   
 {  
 \label{mass-d}
 \includegraphics[scale=0.45]{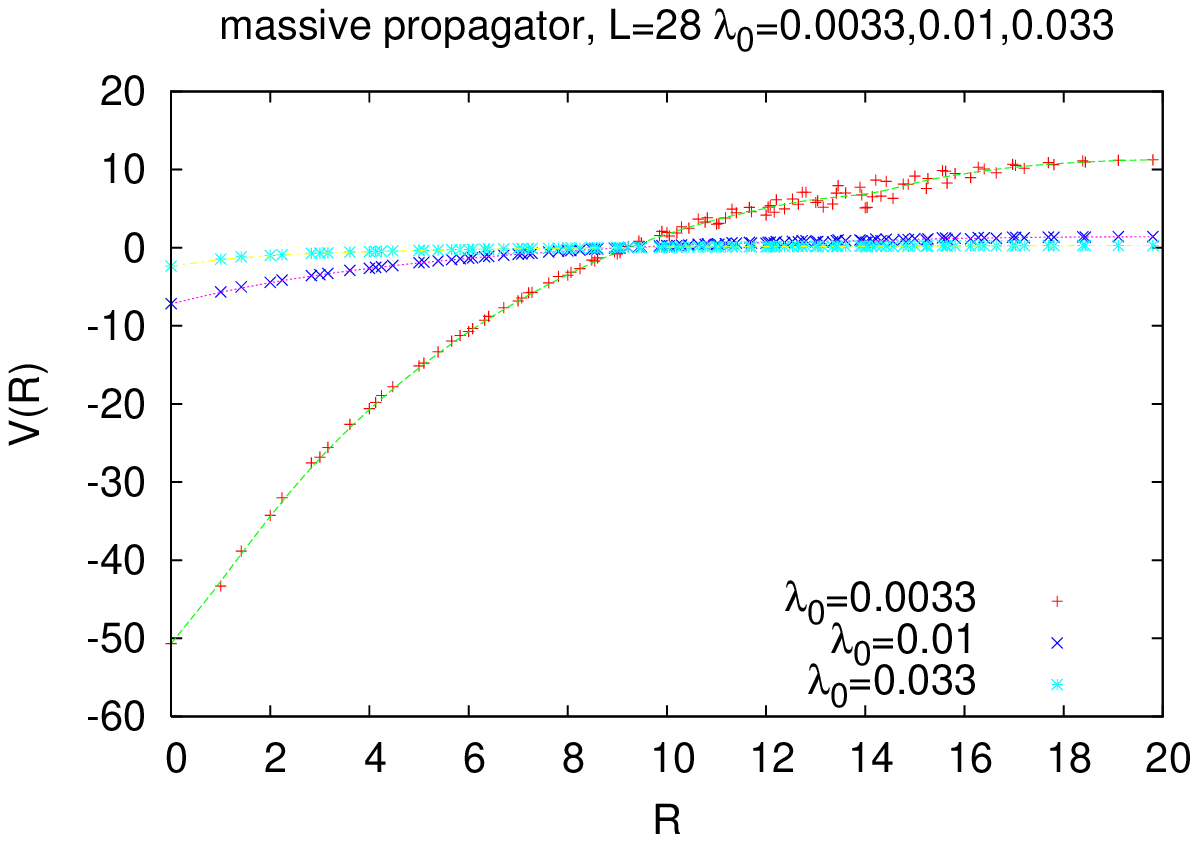}
 }
\end{center}
\caption{Same as previous figure with a massive transverse gluon propagator, and 
$\l_0$ = (a) 0.033, (b) 0.01, (c) 0.0033, (d) combined.}
\label{mass} 
\end{figure}

   The result for three different choices of $\l_0=0.033,0.01,0.0033$ are shown in Fig.\ \ref{gribov}.  There is no clear evidence of a linear potential developing as $\l_0$ decreases.  Instead, the main effect of decreasing $\l_0$ is a drastic
increase in the overall magnitude of the color Coulomb potential.   We can also check whether using a massive transverse gluon propagator, rather than the Gribov form, would make any difference.  In fact
there is no qualitative difference, as seen in Fig.\ \ref{mass}.   

  It was argued in ref.\  \cite{Greensite:2004ur} that confinement calls for not only a small value of $\l_0$, but also an enhanced density of near-zero eigenvalues.  This may not be a feature of the test configurations we have chosen.  It is also evident that the $A^a_i(k)$ components generated by our procedure are completely uncorrelated, e.g.\
$\langle A_i^a(q_1) A_j^b(q_2) A_k^c(q_3) \rangle  =   0$, which is obviously unrealistic in the infrared regime of a non-abelian gauge theory. 

\section{Conclusion}

    It is found that both  the GO and hybrid proposals for the Yang-Mills vacuum wavefunctional, which both have the property of dimensional reduction, fit the data for abelian plane waves, and  non-abelian constant lattices, almost perfectly.  The mass gap and the Coulomb gauge ghost propagator also work out well.  The Coulomb gauge proposal $\Psi_{CG}$ is consistent with the abelian plane wave
measurements, for the choice $c_1=0$.   It fails on non-abelian constant configurations.

   We conclude that the data supports the conjecture of dimensional reduction in the infrared; i.e.\  long wavelength vacuum fluctuations in 2+1 dimensions, at fixed time, resemble fluctuations in a two-dimensional Euclidean theory.   
A second conclusion is that proximity to the Gribov horizon does not, by itself, seem sufficient to produce a confining Coulomb potential, as we have seen in a particular example. Field configurations which do give the confining result must evidently satisfy some other conditions, apart from proximity to the horizon.
    


\begin{thebibliography}{99}
\bibitem{Greensite:2011pj}
  J.~Greensite, H.~Matevosyan, {\v S}. Olejn\'{\i}k, M.~Quandt, H.~Reinhardt, A.~P.~Szczepaniak,
  Phys.\ Rev.\  {\bf D83}, 114509 (2011).
  [arXiv:1102.3941 [hep-lat]].

\bibitem{Greensite:1979yn}
  J.~P.~Greensite,
  Nucl.\ Phys.\  {\bf B158}, 469 (1979).

\bibitem{Greensite:2007ij}
  J.~Greensite, {\v S}. Olejn\'{\i}k,
  Phys.\ Rev.\  {\bf D77}, 065003 (2008).
  [arXiv:0707.2860 [hep-lat]].

\bibitem{Karabali:1998yq}
  D.~Karabali, C.~-j.~Kim, V.~P.~Nair,
  Phys.\ Lett.\  {\bf B434}, 103-109 (1998).
  [hep-th/9804132].

\bibitem{Szczepaniak:2001rg}
  A.~P.~Szczepaniak, E.~S.~Swanson,
  Phys.\ Rev.\  {\bf D65}, 025012 (2002).
  [hep-ph/0107078]; \\
  A.~P.~Szczepaniak,
  Phys.\ Rev.\  {\bf D69}, 074031 (2004).
  [hep-ph/0306030].

\bibitem{Feuchter:2004mk}
  C.~Feuchter, H.~Reinhardt,
  Phys.\ Rev.\  {\bf D70}, 105021 (2004).
  [hep-th/0408236]; \\
  H.~Reinhardt, C.~Feuchter,
  Phys.\ Rev.\  {\bf D71}, 105002 (2005).
  [hep-th/0408237].

  
\bibitem{Greensite:1987rg}
  J.~Greensite,
  Phys.\ Lett.\  {\bf B191}, 431 (1987); \\
  J.~Greensite, J.~Iwasaki,
  Phys.\ Lett.\  {\bf B223}, 207 (1989).

\bibitem{Meyer:2004jc}
  H.~B.~Meyer, M.~J.~Teper,
  Phys.\ Lett.\  {\bf B605}, 344-354 (2005).
  [hep-ph/0409183].

\bibitem{Matevosyan:2008nf}
  H.~H.~Matevosyan, A.~P.~Szczepaniak, P.~O.~Bowman,
  Phys.\ Rev.\  {\bf D78}, 014033 (2008).
  [arXiv:0805.0627 [hep-ph]].

\bibitem{Quandt:2011zz}
  M.~Quandt, G.~Burgio, H.~Reinhardt,
  AIP Conf.\ Proc.\  {\bf 1343}, 206-208 (2011).

\bibitem{Greensite:2004ur}
  J.~Greensite, {\v S}. Olejn\'{\i}k, D.~Zwanziger,
  JHEP {\bf 0505}, 070 (2005).
  [hep-lat/0407032].

\end{thebibliography}
\end{document}